# Insights into the formation mechanism of two-dimensional lead halide nanostructures

Eugen Klein[1], Rostyslav Lesyuk[1,2], Christian Klinke[1,3,*]

[1] *Institute of Physical Chemistry, University of Hamburg,*
*Grindelallee 117, 20146 Hamburg, Germany*
[2] *Pidstryhach Institute for applied problems of mechanics and mathematics of NAS of Ukraine,*
*Naukowa str. 3b, 79060 Lviv, Ukraine*
[3] *Department of Chemistry, Swansea University - Singleton Park, Swansea SA2 8PP, UK*

**Abstract**

*We present a colloidal synthesis strategy for lead halide nanosheets with a thickness of far below 100 nm. Due to the layered structure and the synthesis parameters the crystals of $PbI_2$ are initially composed of many polytypes. We propose a mechanism which gives insight into the chemical process of the $PbI_2$ formation. Further, we found that the crystal structure changes with increasing reaction temperature or by performing the synthesis for longer time periods changing for the final 2H structure. In addition, we demonstrate a route to prepare nanosheets of lead bromide as well as lead chloride in a similar way. Lead halides can be used as a detector material for high-energy photons including gamma and X-rays.*

\* Corresponding author: klinke@chemie.uni-hamburg.de

## 1. Introduction

Two-dimensional nanostructures represent an important field in colloidal syntheses with a wide range of applications.[1,2,3] Iron sulfide ($Fe_3S_4$) nanosheets (NS) for example, are used as anode materials for lithium-ion batteries.[4] Cadmium selenide (CdSe) is one of the best-known semiconductor materials and finds application in many fields of research like in photocatalysis or in optoelectronics.[5,6] Further, various other two-dimensional semiconductor nanocrystals are suited as materials in transistor devices due to the high conductivity in plane.[7,8,9] An important scientific field is represented by materials like gallium nitride (GaN) or gallium oxide ($Ga_2O_3$) which are used as UV photodetectors due to their large band gap.[10,11] Compared to thin films composed of small nanoparticles, 2D nanostructures have the advantage that they do not exhibit tunnel barriers or grain boundaries in the lateral dimensions, which makes them interesting for optoelectronics, photovoltaics[12,13], in particular for flexible electronic devices.[14]

There are many ways to synthesize 2D nanostructures like exfoliation of layered structures, solvothermal methods, chemical vapor deposition, and direct colloidal synthesis in a flask. Chemical exfoliation based techniques use either neutral or charged molecules or ions in order to separate the layered structures through intercalation, yielding single- or multilayer nanosheets.[15,16,17] Solvothermal methods are performed in a stainless steel autoclave in water (hydrothermal) or other polar solvents like ethylene glycol at high pressure.[18,19] Thin sheets prepared by chemical vapor deposition grow on a substrate which was placed in a closed reaction chamber that is filled with one or more volatile precursors. By using this technique it is possible to produce high quality solid materials like GeSe, $MoSe_2$ or $WSe_2$.[20,21,22] The synthesis in a flask can be executed through a single source precursor,[23] by adding one of the reaction partners with a syringe pump over a certain time period,[24] or by using the hot injection method where the second reactant is rapidly introduced to the reaction mixture.[25] An important advantage of the methods following the colloidal route is low-cost solution processability.

In order to obtain nanostructures with specific shapes like 0D spheres, 1D wires or 2D sheets, ligands which passivate energetically favorable facets are the most important factor. Metal-free ligands such as chalcogenides and hydrochalcogenides ($S^{2-}$, $HS^-$, $Se^{2-}$, $HSe^-$, $Te^{2-}$, $HTe^-$, $TeS_3^{2-}$) [26] or metal-containing ligands like a tungsten arsenate oxide[27] are used in various synthesis of nanomaterials, however the most prominent ligands are organic compounds like oleic acid, oleylamine or trioctylphosphine (TOP).[28,29] Additionally, co-ligands namely halogen alkanes can be used to synthesize CdSe pyramids or PbS nanosheets.[13,30]

Lead iodide is a direct band gap semiconductor with a gap between 2.3 and 2.4 eV[31] and a crystal structure which consist of layers of hexagonally close packed iodine and lead atoms oriented perpendicular to the c-axis.[32,33,34] Using specific synthesis parameters it is possible to prepare crystals of $PbI_2$ with specific orientations of the layers. These specific orientations of layers are called polytypes. The most frequent polytype is the 2H lead iodide[34] having the stacking sequence (AαB), where A, B denote iodine ions, and α the lead ion.

The potential applications for this material are high energy photon detectors for X-rays and gamma rays and photocells.[32] $PbBr_2$ and $PbCl_2$ are wide band gaps materials with an orthorhombic crystal structure. All three lead halides find application as precursor materials in the perovskite solar cell fabrication.[35]

Here, we report on the synthesis and characterization of $PbI_2$, $PbBr_2$ and $PbCl_2$ nanosheets prepared via a direct colloidal route. The nanosheets are analyzed by TEM, XRD, AFM and UV/Vis techniques. To our best knowledge we report for the first time syntheses of two-dimensional $PbBr_2$, $PbCl_2$ and $PbI_2$ sheets with thicknesses far below 100 nm. In addition to the possible formation process, we also provide insights into the thermodynamically triggered change of the crystal structure.

## 2. Experimental Section

*Synthesis*

All chemicals were used as received: Lead(II) acetate tri-hydrate (Aldrich, 99.999%), oleic acid (OA, Aldrich, 90%), 1-chlorotetradecane (CTD; Aldrich, 98%), 1-bromotetradecane (BTD; Aldrich, 97%), tri-octylphosphine (TOP; ABCR, 97%), and 1,2-diiodoethane (DIE; Aldrich, 99%).

$PbI_2$ synthesis: In a typical synthesis a three neck 50 mL flask was used with a condenser, septum and thermocouple. 860 mg of lead acetate tri-hydrate (2.3 mmol) were dissolved in 20 mL of oleic acid (60 mmol) and heated to 80 °C until the solution turned clear in a nitrogen atmosphere. Then vacuum was applied to remove the acetic acid which is generated by the reaction of oleic acid with the acetate from the lead precursor. After 1.5 h the reaction apparatus was filled with nitrogen again and 2 mL of a 48.7 mg 1,2-diiodoethane (0.17 mmol) in 3 mL oleic acid precursor was added at 80 °C to the solution. After 4 minutes 0.06 mL of tri-octylphosphine (0.13 mmol) was added to the reaction solution. After 4.5 – 64.5 minutes the heat source was removed and the solution was left to cool down below 60 °C. Afterwards, it was centrifuged at 4000 rpm for 3 minutes. The particles were washed two times in toluene before the product was finally suspended in toluene again and put into a freezer for storage.

$PbBr_2$ synthesis: The procedure was similar to the $PbI_2$ synthesis except of a higher reaction temperature and a different moment for introducing the bromine source. 860 mg of lead acetate tri-hydrate (2.3 mmol) were dissolved in 5 mL of oleic acid (15 mmol) and 10 mL of 1-bromotetradecane (34 mmol) and heated to 75 °C until the solution turned clear in a nitrogen atmosphere. The reaction was started by adding 0.06 mL of tri-octylphosphine (0.13 mmol) at a temperature of 150 °C and was stopped 11 minutes later.

$PbCl_2$ synthesis: The procedure was similar to the $PbI_2$ and $PbBr_2$ synthesis with just a different amount of the chloride source and a higher reaction temperature. 860 mg of lead acetate tri-hydrate (2.3 mmol) were dissolved in 3.5 mL of oleic acid (10.5 mmol) and 15 mL of 1-chlorotetradecane (54 mmol) and heated to 75 °C until the solution turned clear in a nitrogen atmosphere. The reaction was started by adding 0.06 mL of tri-octylphosphine (0.13 mmol) at a temperature of 180 °C and stopped 10 minutes later.

Investigation on the reaction mechanism: To ensure the presence of the 1,2-diiodoethane peak in the $^1H$ NMR the synthesis for the $PbI_2$ nanosheets was performed at higher concentrations.

564 mg of 1,2-diiodoethane (2 mmol) were dissolved in 4 mL of oleic acid and 2.66 mL of this precursor were added to the reaction mixture described above.

*TEM*

The TEM samples were prepared by diluting the nanosheet suspension with toluene followed by drop casting 10 µL of the suspension on a TEM copper grid coated with a carbon film. Standard images were done on a JEOL-1011 with a thermal emitter operated at an acceleration voltage of 100 kV.

*XRD*

X-ray diffraction measurements were performed on a Philips X'Pert System with a Bragg-Brentano geometry and a copper anode with a X-ray wavelength of 0.154 nm. The samples were measured by drop-casting the suspended nanosheets on a <911> or <711> cut silicon substrate.

*AFM*

Atomic force microscopy measurements were performed in tapping mode on a Veeco MultiMode NanoScope 3A and a JPK Nano Wizard 3 AFM in contact mode. The samples were prepared by drop-casting a diluted nanosheet suspension on a silicon wafer.

*Spectroscopy*

UV/vis absorption spectra were obtained with a Cary 5000 spectrophotometer equipped with an integration-sphere. The PL spectra measurements were obtained by a fluorescence spectrometer (Fluoromax-4, Horiba Jobin Yvon).

*Acknowledgments*

The authors thank the German Research Foundation DFG financial support in the frame of the Cluster of Excellence "Center of ultrafast imaging CUI" and for granting the project KL 1453/9-2. The European Research Council is acknowledged for funding an ERC Starting Grant (Project: 2D-SYNETRA (304980), Seventh Framework Program FP7).

## 3. Results and Discussion

**Chemical reaction.** Controlling the size and shape of nanocrystals requires the knowledge of the function and purpose of every reactant which participates in the reaction. Therefore a fundamental investigation and understanding of the mechanism is crucial. The two-dimensional $PbI_2$ NS are produced by injecting a DIE oleic acid solution and TOP separately into a degassed lead oleate oleic acid mixture preheated to 80 °C. In order to verify the exact function of TOP and any other reactions in the synthesis several aliquots taken during the reaction were investigated by $^1H$ and $^{31}P$ NMR spectroscopy. Figure 1A shows the $^1H$ NMR spectrum of aliquots taken from one and the same synthesis at different times. All of the peaks in the spectra belong to oleic acid except the one at 2.67 ppm and nearly all of them show no shift as a function of changes in the environment like pH or volume. The peak at 2.67 ppm can be assigned to 1,2-diiodoethane with a singlet due to the equivalence of the protons (Figure 1B). The only peak that is shifting is the proton of the carboxylic group between 12.25 ppm and 12.32 ppm shown in Figure 1C. Further, $^{31}P$ NMR was performed for the last step in which TOP was added. Figure 1D shows that even right after the injection all of the TOP molecules have already reacted and only TOPO can be observed. The TOPO shift appears at slightly higher values due to a change in the local environment with the formation of $PbI_2$ nanoparticles.[36] Based on these results we propose a mechanism depicted in Scheme 1 for the synthesis of $PbI_2$ nanoparticles which takes place in two steps. The first step starts after the injection of 1,2-diiodoethane which reacts with oleic acid resulting in the substitution of one of the iodide ions. The free iodide ion and the proton of the coordinated oleic acid forms hydroiodic acid and thereby decreases the pH-level of the solution. This change can be followed by the shift of the peak of the carboxylic group from 12.25 ppm to 12.27 ppm. At last the hydroiodic acid reacts with the lead oleate to form $PbI_2$ monomers and oleic acid. Since the time duration between the first and second aliquot was 4 min but the shift of the peak was small as well as no color change could be observed in the reaction mixture we believe that this first step happens very slowly. Syntheses without TOP at 200 °C and above lead to $PbI_2$ sheets. Therefore, TOP is not needed for the reaction to take place (Figure S1 A). A more direct approach to prepare $PbI_2$ nanosheets were syntheses with potassium iodide (KI) and ammonium iodide ($NH_4I$). Particles prepared by following these procedures could only be obtained at temperatures of 150 °C and above and showed thicknesses similar to the NS synthesized without TOP which were higher than 50 nm.

The second step starts with the injection of TOP which substitutes an iodide atom at the 1,2-diiodoethane and resulting in the release of an iodide ion. This reacts again to hydroiodic acid and further with lead oleate to a lead oleate iodide complex which can be observed by the shift of the carboxylic group from 12.27 ppm to 12.31 ppm. Finally the lead oleate iodide complex can react in the way proposed in Scheme 1 B following step 1 or step 2 to form $PbI_2$ monomers and a TOP oleate molecule. This phosphor molecule forms an anhydrate and TOPO that can be detected in the $^{31}P$ NMR.[36] Due to the relatively strong shift from 12.27 ppm to 12.31 ppm right after the injection of TOP we believe that a large amount of hydroiodic acid is produced and therefore the second step occurs much faster than the first one. Further indications are provided by the $^{31}P$ NMR taken after the TOP injection. The spectrum shows only the peak for TOPO and that the reaction mixture turns from clear colorless to a turbid yellow-green solution. This means that all TOP molecules react instantly with 1,2-diiodoethane, producing a large amount of $PbI_2$ monomers. This process leads to a much faster supersaturation of the mixture and therefore an increase of the reactivity in this system. The last aliquot shows a shift of the carboxylic proton from 12.31 ppm to 12.29 ppm. 4 min after the injection of TOP most of the hydroiodic acid produced at the beginning of the second step was used up and the pH has increased. The fact that the peak for the 1,2-diiodoethane appears in the last $^1H$ NMR spectrum means the reaction is not finished. The reaction parameters were selected to produce sheets with the smallest thickness. Therefore, an amount of lead and iodide precursor was chosen which could not completely react within the given time period. Many authors have reported about the effect of phosphines and phosphine impurities to reduce lead oleate to $Pb^0$ species.[36,37,38] In our case the formation of $Pb^0$ can be excluded due to several reasons. Literature reports describe reactions at temperatures above 140 °C (and some state that) with a large amount of TOP (5 mL) and reaction times of hours are needed to generate $Pb^0$. In contrast to them, our reaction temperature lies at 80 °C and the amount of TOP is very limited (0.06 mL) for the described mechanism. Moreover there is no chemical in our approach which could oxidize $Pb^0$ back to the $Pb^{2+}$ species.

**Change of the space group.** Figure 2 shows representative TEM images for $PbI_2$ nanosheets prepared by varying the growth time between 0 min and 60 min with all other reaction parameters remaining constant. All four samples consist of nanosheets with a hexagonal shape while exhibiting a strong tendency to stack. The dimensions of these particles are more or less the same having lateral sizes between 1.5 µm and 4 µm. On the basis of the data shown in Figure 2 the sheets have a homogeneous surface and well defined edges. Figure S1 B presents an overview for particles prepared immediately after the formation. The thickness was calculated by measuring the width of the reflex at 39 ° in the powder X-ray diffractograms (Figure 3). This reflex represents the <001> direction in our XRD measurements on the substrate. By using a Gauss fit for the reflex at 39 ° the full width at half maximum (FWHM) changes only slightly with longer growth periods. The calculated thicknesses from the FWHM values using the Scherrer equation[39] (with a form factor of 1) were 21.3 nm (0 min), 18.7 nm (5 min), 17.3 nm (20 min), and 27.6 nm (60 min) respectively. The X-ray diffractogram of the sheets prepared immediately after the formation reveal a hexagonal crystal structure with the P3m1 space group (Figure 3). Most of the theoretically possible reflexes do not appear due to the planar orientation of the sheets on the substrate. The tendency of the sheets to orientate themselves parallel on the substrate can be described as a texture effect where only lattice planes parallel to the substrate can be measured in the possible angle range. Increasing the growth time to 5 min and 20 min gives rise to some additional reflexes at the angles of 12 °, 25 ° and 52 ° while reducing the intensity of the peaks at 15 °, 23 °, 46 ° and 56 °. Syntheses stopped after even longer growth times like 1 h show the complete omission of the peaks at 15 °, 23 °, 46 ° and 56 °. This indicates that the space group of the particles shifts from the P3m1 to the more symmetric P-3m1 with time. Based on the missing of a large number of reflexes in all the shown XRDs and therefore to ensure the crystal structure, X-ray diffraction in a capillary was performed on the products prepared after 0 min and 60 min (Figure S2). For the sheets synthesized after 60 min all of the corresponding reflexes can be observed while the diffractogram for the 0 min sample does not show all of the reflexes for the P3m1 space group. A reason for this is probably the relatively small intensities of the major part of the reflexes of 0.1 % taken from the literature which makes them disappear in the noise of the measurement. More important is the fact that the particles prepared after 60 min have only the reflexes of the space group P-3m1 thus the missing of these reflexes was not due to the texture effect. In an attempt to assign the right polytype to the sample obtained immediately after the sheets formation, we compared the measured capillary XRD pattern with simulated XRD patterns for 2H, 4H, 6H, 8H, 10H, 14H and 20H structures. For the 4H, 10H and 20H polytypes the crystallographic data are based on references [34,40]. For other polytypes higher than 4H we built similar sequences always adding another (AαB) block to the structure. For the 20H the sequence (AαB)(AαB)(AαB)(AαB)(AαB)(AαB)(AαB)(AαB)(CβB)(CβB) was adapted from Ref. 40. Our study indicates that the kinetic product of our synthesis after 0 min may contain several polytypes,[41,42] including 10H, 20H and the 4H (Figure S3 and further discussion in the Supporting Information).

A Schematic illustration of the process occurring during and after the formation of the nanosheets with growth time for the 10H polytype is given in the Figure 4. At first, the more asymmetric space group P3m1 is formed which has a large lattice parameter (for the (100) plane of 3.4895 nm, for the 20H c=6.979 nm respectively) and one displaced layer of lead and iodide, respectively. With longer growth periods the displaced planes shift in order to form a more symmetric crystal structure. At this point the reflexes of the space group P-3m1 appear while at the same time some reflexes of the P3m1 space group start to disappear. At the end of this process, meaning after one hour growth time the more symmetric space group P-3m1 remains to define the two-dimensional crystals. Thus the polytypic transition occurs.

**Comparison with PbBr$_2$ and PbCl$_2$.** The PbX$_2$ (X= I, Br, Cl) nanosheet syntheses have a few things in common: the same amount of lead acetate tri-hydrate and TOP. Latter was used in all cases in order to increase the reactivity. For the PbI$_2$ synthesis oleic acid is used in excess and serves as solvent and ligand. The iodide source is 1,2-diiodoethane which is dissolved in oleic acid and must be prepared one to two hours before the injection since it is not long-term stable in solution even when stored in the fridge. TOP is essential for this method due to the fact that the reactivity without is not high enough and no material would be formed at the given temperatures. The so prepared nanosheets exhibit a uniform size and shape having lateral dimensions of 1.5 µm to 4 µm (Figure 5 A). Nevertheless, the size of the sheets can be controlled between 2 µm and 15 µm by varying the temperature (Figure S4). The electron diffraction pattern in Figure 5 B taken from a single sheet shows a dot pattern and indicates that the prepared nanosheets are single crystals. The syntheses of PbBr$_2$ and PbCl$_2$ are carried out in an excess of the corresponding halide sources and at higher temperatures compared to the PbI$_2$ synthesis. By using an excess of long chained haloalkanes like 1-bromotetradecane and 1-chlorotetradecane the reactivity of the reaction can be controlled to a certain degree. Performing the syntheses with haloalkanes possessing shorter alkyl chains like dibromoethene and dichloroethene leads to structures which are several hundred nanometers thick. Higher temperatures of 150 °C and 180°C are necessary to generate bromine ions and chlorine ions. Figure 5 C shows a TEM image of PbBr$_2$ nanosheets which have a smooth surface and a hexagonal shape. These particles as is the case with PbI$_2$ show a tendency to stack and exhibit a large size distribution between 1 µm and 4 µm. The PbCl$_2$ structures have a stripe- or rod-like shape with lengths between 2 µm to 4 µm and widths between 50 nm to 700 nm (Figure 5 D). As mentioned in the introduction these materials can be used as precursors for the preparation of the corresponding perovskite structures. Figure S5 shows TEM images, selected area diffraction pattern (SAED) and XRD of as prepared PbI$_2$ methylammonium iodide particles. They exhibit a hexagon like shape while the single crystal of the PbI$_2$ sheets is somewhat damaged. The reflexes in the XRD fit well with the literature for the perovskite material.

To discuss the thickness and crystal structure of the structures powder X-ray diffraction as well as atomic force microscopy measurements were carried out. Figure S6 shows powder XRDs of $PbI_2$, $PbBr_2$ and $PbCl_2$. $PbI_2$ has a hexagonal crystal structure with a space group of P-3m1 while $PbBr_2$ and $PbCl_2$ have an orthorhombic structure with the Pnam space group. The omission of most of the peaks for all three materials can be observed and occurs as a result of the texture effect described earlier. The AFM images and height profiles of $PbI_2$, $PbBr_2$ and $PbCl_2$ are shown in Figure S7. The surface of $PbI_2$ NS depicted in Figure S7 A is not flat compared with $PbBr_2$ nanosheets (Figure S7 B). The calculated thickness of $PbI_2$ from the XRD data was 21.3 nm and from the AFM measurement 25 nm including the oleic acid shell of 3.6 nm.[43] However, the thickness can be varied by many parameters like the temperature, the Pb:I ratio or nature of the solvent (Figure S8). In contrast to $PbI_2$, $PbBr_2$ and $PbCl_2$ show large differences in the values obtained from XRD and AFM. The thicknesses calculated from XRD were 56 nm and 90 nm and from AFM 20 nm and 63 nm for $PbBr_2$ and $PbCl_2$. The larger thicknesses for these two materials are obtained due to the higher reaction temperatures for $PbBr_2$ at 150 °C and $PbCl_2$ at 180 °C compared to $PbI_2$ at 80 °C. The reason for the difference in thickness between AFM and XRD can already be observed in the TEM images for $PbBr_2$ and $PbCl_2$ in Figure 5 where the darker particles are not solely stacked sheets but also thicker structures. We note that despite the identical crystal structure, the $PbBr_2$ and $PbCl_2$ nanosheets have different shape. $PbBr_2$ tends to form hexagonal structures, the $PbCl_2$ forms elongated 2D structures. The reason behind this peculiarity is the different basal planes of the nanosheets apparently due to different synthesis conditions and unit cell constants. From the XRD and SAED data we conclude that for the $PbBr_2$ the basal plane is the (010) crystallographic plane. In XRD data we observe only the reflexes (020), (040), (060) due to texture effects (Fig. S6 A). The $PbCl_2$ nanosheets crystallize and grow parallel to the (120) plane (we observe the (120), (240) and the (360) reflexes, Fig. S6 A). These data are supported by the SAED patterns (Fig S6 B,C). This in turn causes different growth directions and shape of the synthesized 2D structures. The shape of the $PbI_2$ structures which have hexagonal symmetry with the basal plane (001) is hexagonal as expected.

In order to compare the optical properties of lead halide nanosheets, UV/Vis absorbance and PL spectroscopy were carried out. In the case of $PbI_2$, the measurements supported our findings about the structural changes in nanosheets during the synthesis route. Figure 6 A illustrates the UV/Vis spectral absorption and PL of $PbI_2$ nanosheets extracted during the first minutes after formation. The absorbance shows the presence of peaks at around 420 nm and 497 nm and PL at 469 nm and 562 nm. The strong peaks at shorter wavelengths are blue shifted compared to the bulk material and belong to 22 nm thick nanosheets. The exciton Bohr radius for $PbI_2$ is 1.9 nm.[44] Therefore, the thickness of the nanosheets cannot be the reason for the observed blue shift. To check if the observed shift could be attributed to the space group change and the polytype crossover, we performed DFT calculations of the band structure for each of the polytypes from 4 to 20H (see Supporting Information). We found, that the bandgap value decreases for higher order polytypes compared to the bulk 2H polytype and thus cannot explain the observed blue-shift (Figure S9 A). Examination of the PL and UV/Vis absorption spectra indicates an excitonic character of the nanosheets prepared right after the formation. In contrast to this, the absorption spectrum of nanosheets after 1 h synthesis is inherent to bulk $PbI_2$, having one absorption shoulder at 500 nm (Figure 6 D), which corresponds to the bulk energy gap value (see Tauc plots in the Supporting Information). The PL spectrum (Figure 6 D) has the main maximum at around 527 nm and residual blue-shifted intensity between 400 nm and 500 nm. The shift in PL therefore is calculated to 297 meV (12.6 % relatively to the 2H structure). Based on these data, we assume that the blue-shift and excitonic character of the sheets in the spectra may come from charge carriers confined in the multilayered domain structure formed at the very beginning of the colloidal synthesis. We simulated this model confining different cells (10H, 8H, 6H, 4H and 2H) within a certain space of vacuum (vacancies) and calculated the relative energy gap increase. As can be seen from the Figure S9 B, the relative blue-shift for the 4H structure in confinement is very close to the shift, that we observe in PL (10.7% vs 12.6%). The optical shift observed in UV/Vis (~430 meV) is nearly 17% (from the Tauc plots – 13.8%, Fig. S10) and may be increased compared to the PL shift due to the self-doping of $PbI_2$ and the Moss-Burnstein effect.[45,46] Thus, we conclude, that confinement is presumably responsible for the 4H polytype formed initially (first seconds after sheet formation) in extent compared to other polytypes. The disorder of the structure is high in the first minutes of the synthesis. This could be also seen in TEM images, the lateral structure of NSs is much less homogeneous compared to the structure of NSs after 1 h with 2H structure (Figure 2 A and D).

The possible reason for the existence of such structure in $PbI_2$ nanosheets may be the fast formation mechanism which leads to 20 nm thick nanosheets, presumably accompanied with planar defect formation and vacancy agglomeration. With larger reaction periods this complex structure relaxes gradually due to ion migration to lower energy positions in the lattice. These gradual changes can be observed both by XRD and PL/UV-Vis spectroscopy. It is worth noting, that interpolytypic transitions for $PbI_2$ were observed and described by other authors, e. g. Minagawa *et al.* observed the 4H to 2H transition in $PbI_2$ bulk crystals at room temperature.[47] Sengupta with co-authors observed a photodegradation process of $PbI_2$ nanoparticles accompanied by spectral changes in UV-Vis absorption. They assumed the formation of multilayered particles and the existence of energy barriers between the layers.[48]

PbBr$_2$ nanosheets show one peak for the absorbance at 335 nm and one peak for the PL at 371 nm which corresponds well with the band gap at 3.3 eV (Figure 6 B).[49] Figure 6 C exhibits the optical results for 2D PbCl$_2$ with the corresponding peaks at 291 nm and 305 nm which also fit well for the band gap at 3.8 eV.[50] The structures for PbBr$_2$ and PbCl$_2$ have large thicknesses and as a result of that no confinement can be observed.

## 4. Conclusion

Until now, it was only possible to synthesize particles of lead iodide with thicknesses of 100 nm and larger. Here, we demonstrated that it is possible to prepare this material with thicknesses below 100 nm and furthermore to synthesize structures of lead bromide and lead chloride in this range. We also present a chemical mechanism for the formation of $PbI_2$ nanosheets. The two important points are the usage of haloalkanes as source materials for the anions and TOP which increases the reactivity of the synthesis. A fast reaction of the TOP molecules with the haloalkanes leads to the release of anions which can react with the lead oleate complex to form lead halide particles. The shape is formed due to the high amount of oleic acid present in all lead halide syntheses. Insights into structural changes of $PbI_2$ nanosheets during the synthesis are given. We can conclude that the space group change from P3m1 to P-3m1 is conditioned by the polytypic transition accompanied with relaxation of the initial defect structure. This influences the spectral properties of synthesized nanosheets and can serve as useful tool for quality control of synthesized 2D materials. The defect and bulk nanosheets can be distinguished by all analytic methods like TEM, XRD, AFM and UV/Vis absorbance and PL spectroscopy as well as DFT simulations. Finally, we extend the 2D family of $PbX_2$ materials presenting the synthesis protocols for $PbBr_2$ and $PbCl_2$ nanosheets with micron-range lateral size. The structures might be interesting for low-cost applications in the field of high-energy detectors or they might serve as base materials for perovskite syntheses.

**Figures**

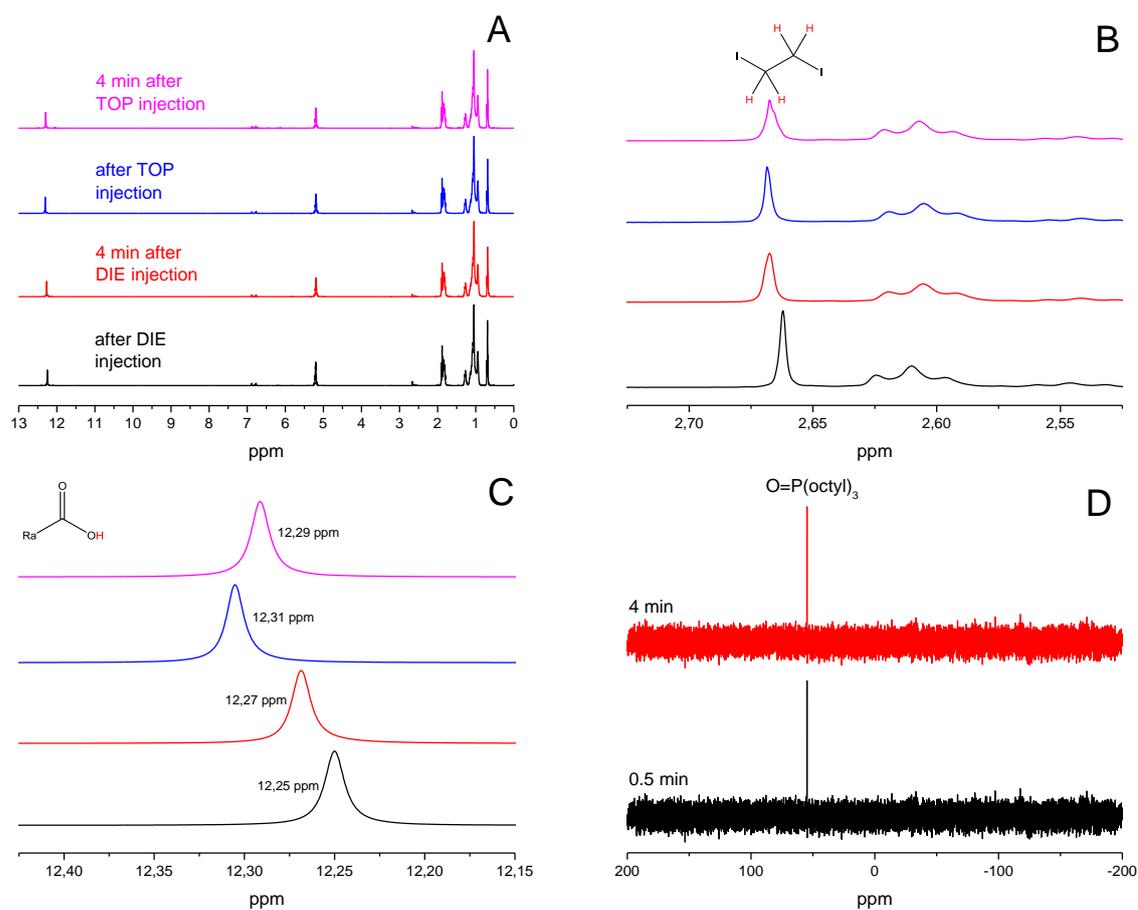

**Figure 1.** $^1$H and $^{31}$P NMR spectra of aliquots taken from one synthesis over time. (A) $^1$H NMR spectra of the four aliquots with all their corresponding peaks. (B) Region of 1,2-diiodoethane peak. (C) Region of the carboxylic group peak and its shifting. (D) $^{31}$P NMR for the aliquots taken after the injection of TOP.

**Scheme 1.** Proposed mechanism for the synthesis of PbI$_2$ nanosheets in two steps. (A) describes the slow reaction where TOP is not involved. (B) shows the synthesis after the injection of TOP.

A

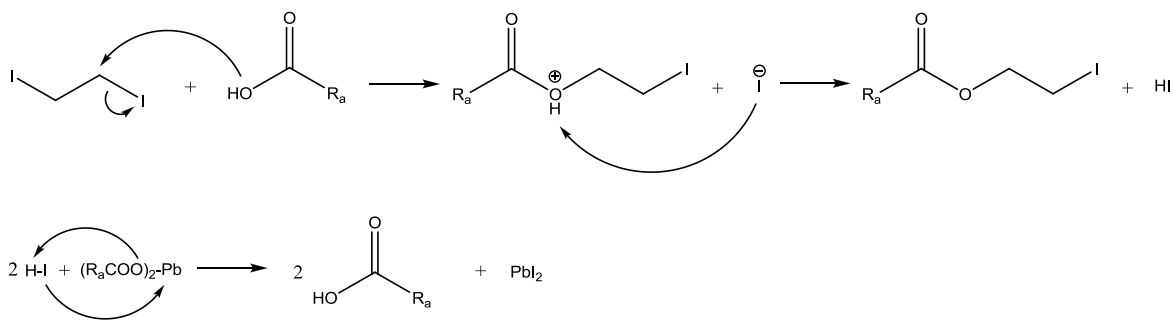

B

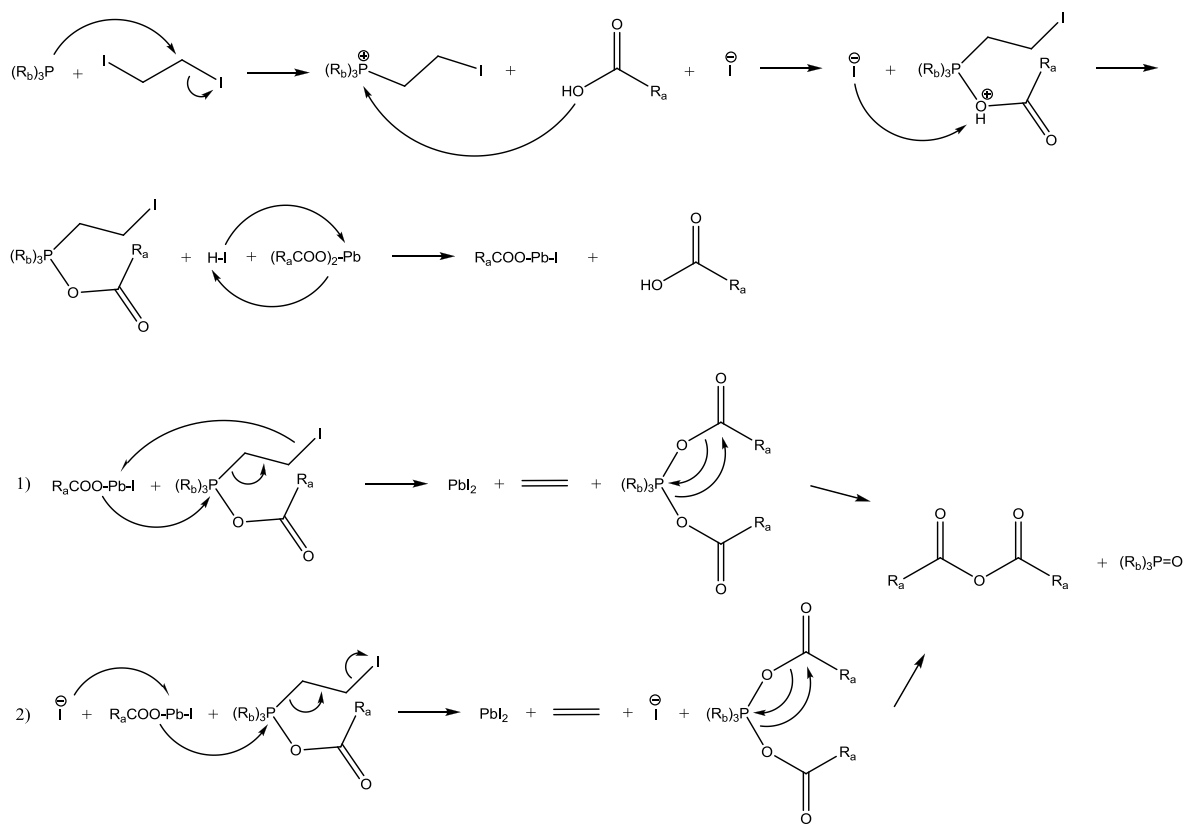

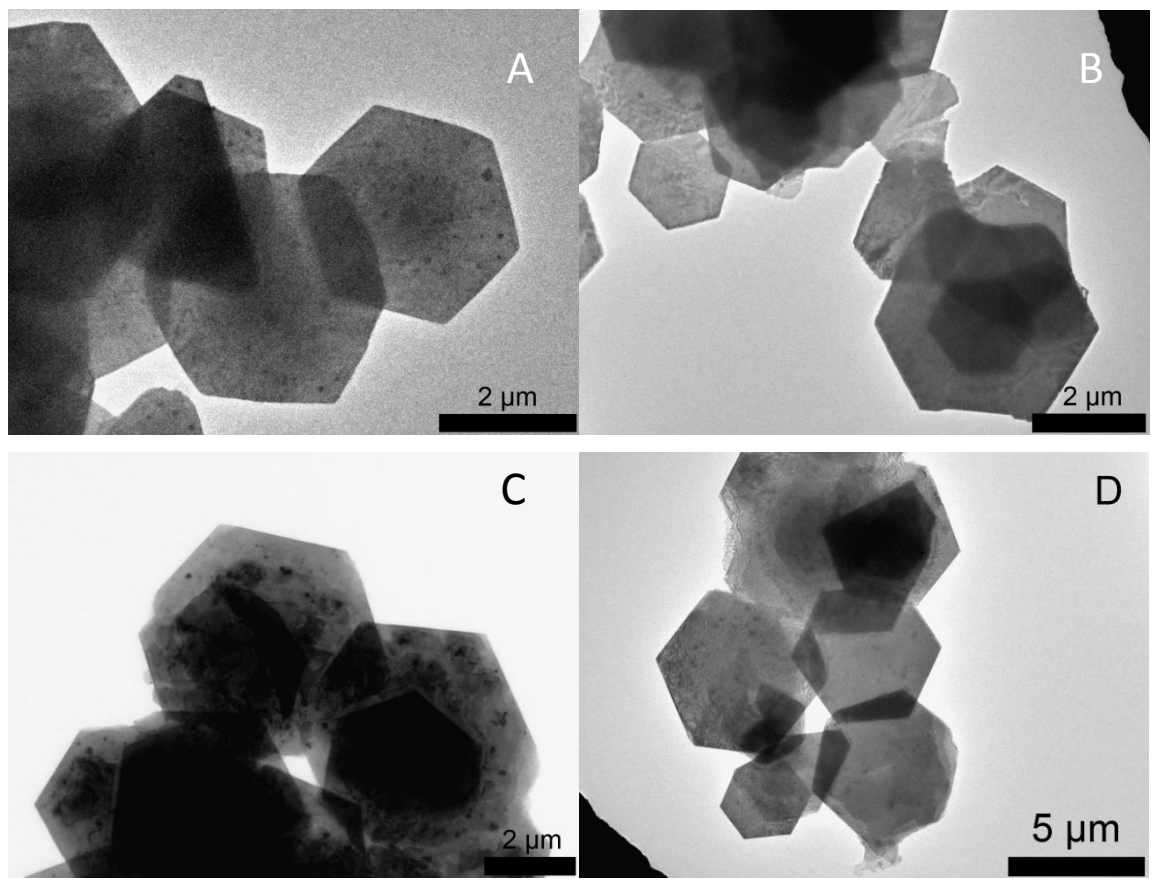

**Figure 2.** TEM images of PbI$_2$ nanosheets prepared by varying the growth time. The growth time increases from immediately after formation (A) to 5 min (B) to 20 min (C), and to 1 h (D) respectively. Despite being prepared for different growth periods, the particles exhibit the same sizes in all three directions. PbI2 is easily affected by the electron-beam and therefore the sheets get slowly destroyed in the TEM. The small dots in Figure 2C are decomposed and remerged material from the sheet structure.

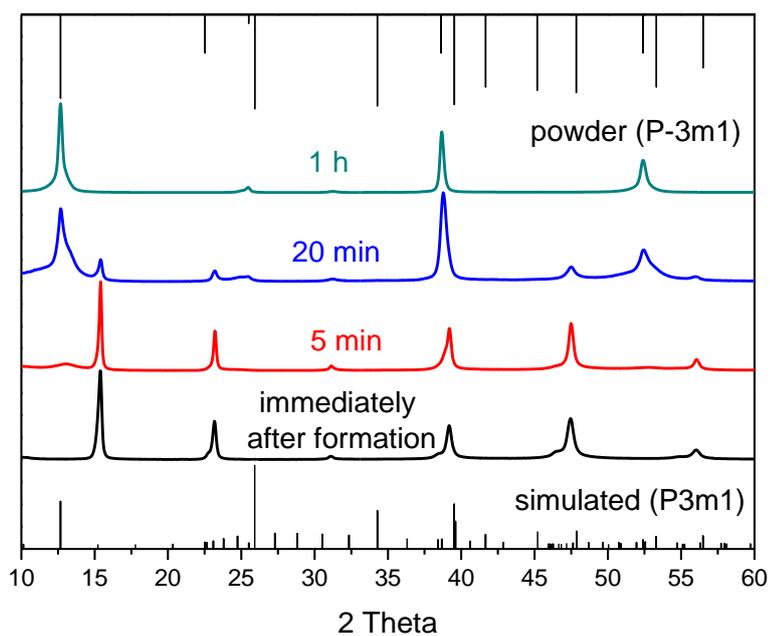

**Figure 3.** Powder XRD of PbI$_2$ nanosheets prepared with varying the growth period. Particles obtained immediately after the formation of the sheets exhibit signals from the P3m1 space group. By extending the time after the formation the crystal structure shifts to the more symmetric space group of P-3m1.

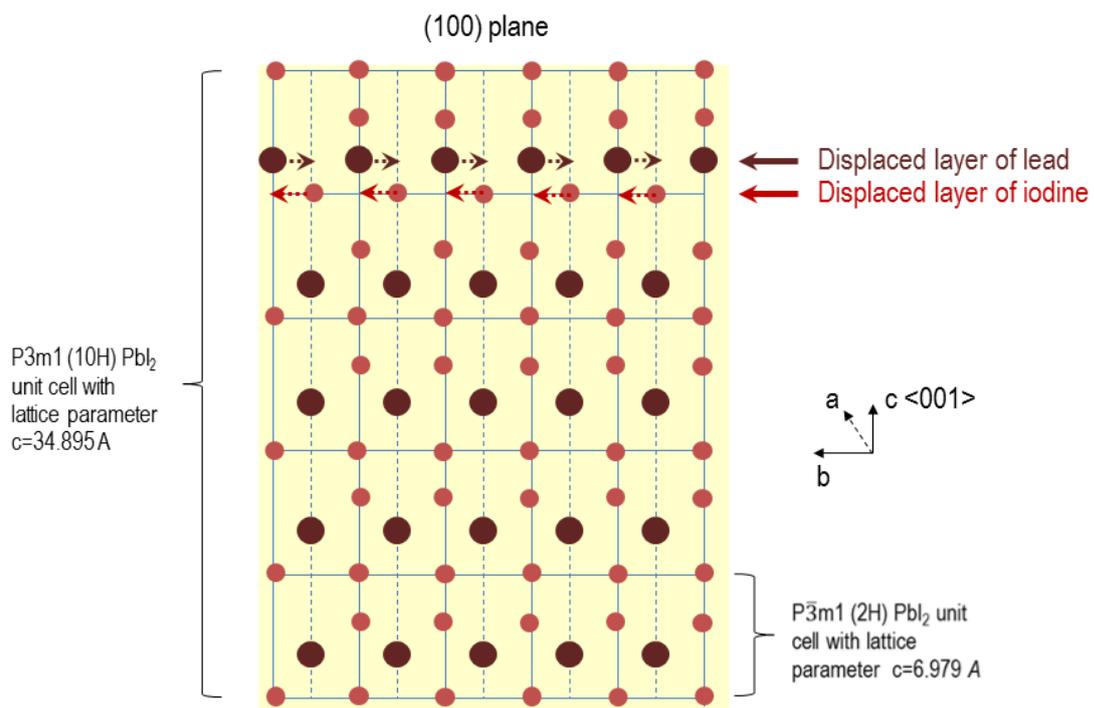

**Figure 4.** Schematic illustration of the change in space group from P3m1 (shown for the 10H polytype) to P-3m1 for PbI2 nanosheets with the growth time.

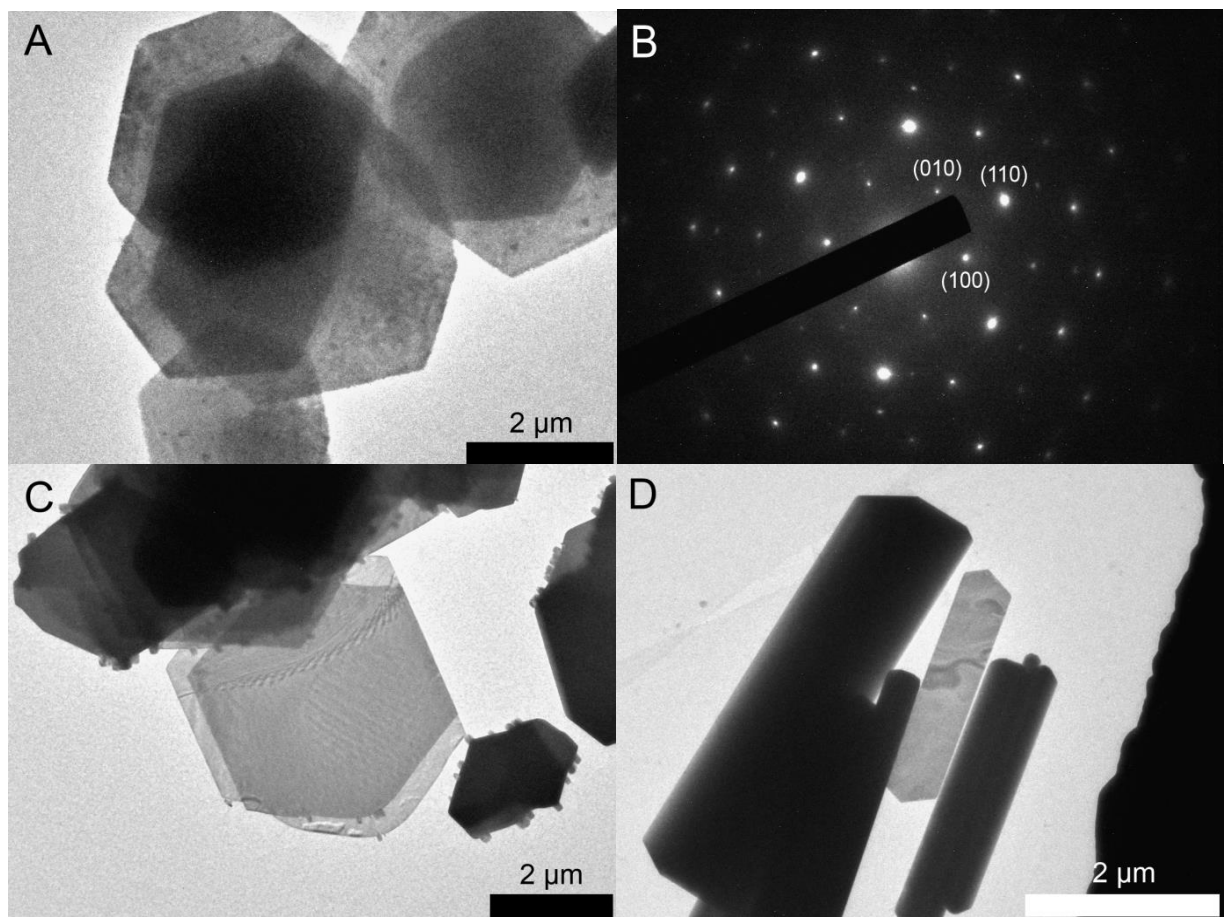

**Figure 5.** TEM images of PbI$_2$, PbBr$_2$ and PbCl$_2$ nanosheets. (A) PbI$_2$ nanosheets with a hexagon-like morphology and (B) electron diffraction pattern of a single PbI$_2$ nanosheet ([001] zone axis). (C) PbBr$_2$ nanosheets with a high size distribution. (D) Partly 2D PbCl$_2$ nanocrystals.

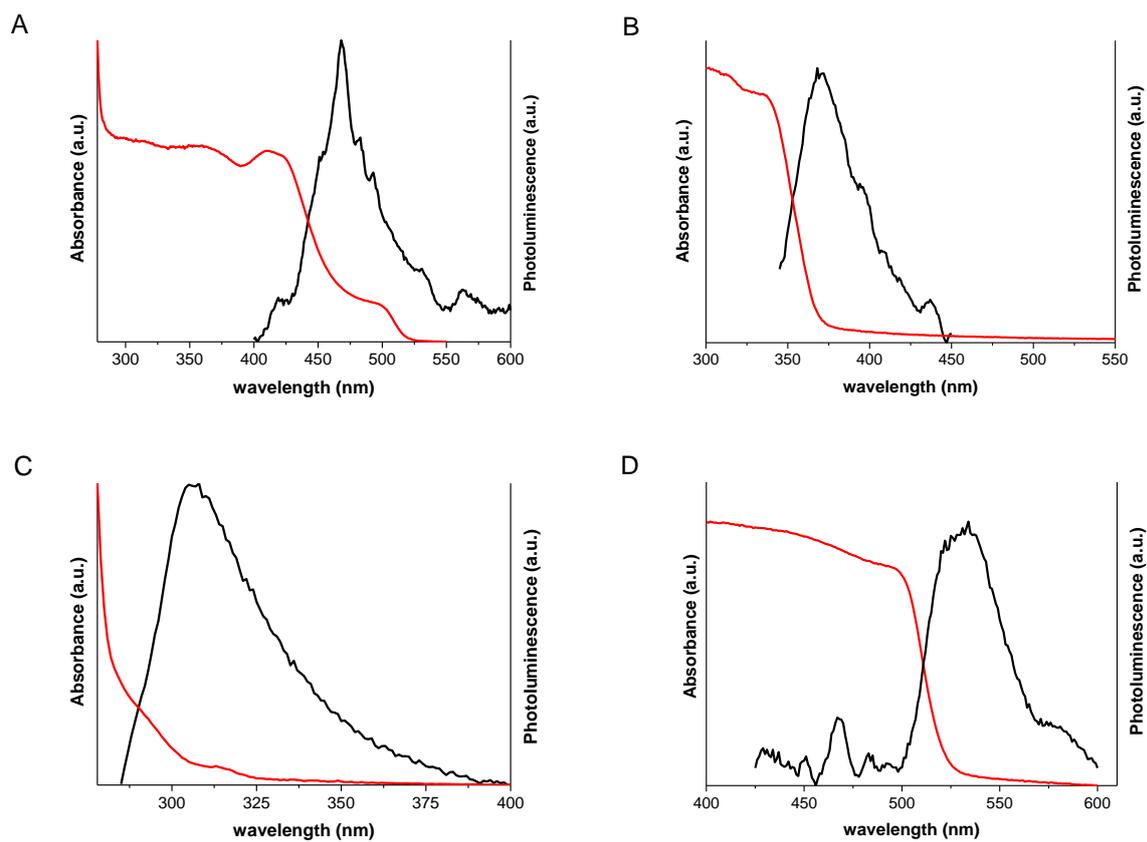

**Figure 6.** Emission and absorbance spectra of the lead halides. (A), PbI$_2$-P3m1; (B), PbBr$_2$; (C), PbCl$_2$; (D), PbI$_2$-P-3m1.